\documentclass[letterpaper, 10 pt, conference]{ieeeconf}  

\IEEEoverridecommandlockouts                              

\usepackage{tikz,pgfplots, pgfplotstable}
\usepgfplotslibrary{statistics}
\usetikzlibrary{patterns}
\usetikzlibrary{matrix}
\pgfkeys{/pgf/number format/.cd,fixed,precision=3}
\pgfplotsset{compat=newest}

\usepackage{nomencl}
\makenomenclature
\usepackage{standalone}
\usepackage{bm}
\usepackage{color}
\usepackage{mathtools,amsmath,amssymb}
\usepackage{algorithm}
\usepackage{algpseudocode}
\usepackage{graphicx}
\usepackage{subfig}
\usepackage{cite}	

\makeatletter
\let\NAT@parse\undefined
\makeatother
\usepackage[hidelinks]{hyperref}
\hypersetup{
	colorlinks,
	citecolor=blue,
	filecolor=black,
	linkcolor=blue,
	urlcolor=blue}

\title{\LARGE \bf
Keeping Them Honest: a Trustless Multi-Agent Algorithm to Validate Transactions Cleared on Blockchain using Physical Sensors*
}

\author{Nikhil Ravi$^{1}$, Shammya Saha$^{1}$, Anna Scaglione$^{1}$ and Nathan G. Johnson$^{2}$
\thanks{*\color{black}This work is supported by Office of Naval Research under the award number N00014-18-1-2393.}
\thanks{$^{1}$Nikhil Ravi, Shammya Saha, and Anna Scaglione are with the School of Electrical, Computer and Energy Engineering, Arizona State University, Tempe, AZ, USA {\tt\small \{nikhil.ravi, shammya.saha, anna.scaglione\}@asu.edu}}%
\thanks{$^{2}$Nathan Johnson is with The Polytechnic School, Arizona State University, Mesa, AZ, USA.
        {\tt\small nathanjohnson@asu.edu}}%
}

\begin{document}

\maketitle

\begin{abstract}

In recent years, many Blockchain based frameworks for transacting commodities on a congestible network have been proposed. In particular, as the number of controllable grid connected assets increases, there is a need for a decentralized, coupled economic and control mechanism to dynamically balance the entire electric grid. Blockchain based Transactive Energy (TE) systems have gained significant momentum as an approach to sustain the reliability and security of the power grid in order to support the flexibility of electricity demand. What is lacking in these designs, however, is a mechanism that physically verifies all the energy transactions, to keep the various inherently selfish players honest. In this paper, we introduce a secure peer-to-peer network mechanism for the physical validation of economic transactions cleared over a distributed ledger. The framework is \textit{secure} in the sense that selfish and malicious agents that are trying to inject false data into the network are prevented from adversely affecting the optimal functionality of the verification process by detecting and isolating them from the communication network. Preliminary simulations focusing on TE show the workings of this framework. 

\end{abstract}


\nomenclature{$T$}{$\in\mathbb{N}$, length of modeling horizon}
\nomenclature{$\mathcal{N}$}{Set of \textit{aggregator regions}}
\nomenclature{$\mathcal{B}$}{Set of buses $b$ in the system}
\nomenclature{$\mathcal{B}_n$}{Set of buses $b$ in aggregator system $n$}
\nomenclature{$\mathcal{E}_e$}{Set of transmission lines $(b,b')$ between buses $b$ and $b'$}
\nomenclature{$\mathcal{E}_c$}{Set of communication lines $(n,n')$ between aggregators $n$ and $n'$}
\nomenclature{$\mathcal{G}_e = (\mathcal{B}, \mathcal{E}_e)$}{Electric grid graph}
\nomenclature{$\mathcal{G}_c = (\mathcal{N}, \mathcal{E}_c)$}{Communication network graph}
\nomenclature{$p_b$}{Active power injection at a bus $b \in \mathcal{B}$}
\nomenclature{$q_b$}{Reactive power injection at a bus $b \in \mathcal{B}$}
\nomenclature{$P_\ell$}{Active power flow over a transmission line $\ell=(b,b') \in \mathcal{E}_e$}
\nomenclature{$Q_\ell$}{Active power flow over a transmission line $\ell=(b,b') \in \mathcal{E}_e$}
\nomenclature{$v_b$}{Voltage magnitude at a bus $b \in \mathcal{B}$}
\nomenclature{$c_\ell$}{Voltage magnitude over a transmission line $\ell=(b,b') \in \mathcal{E}_e$}


\section{Introduction}
\label{sec:intro}
Stressors on the demand and supply chain and congestion in critical infrastructures  like transportation networks (e.g., rails, highways, airports, sea-ports),   telecommunication networks (by frequency-bounded airwaves or cables), and utilities (e.g., electric power, water, gas, oil, sewage) are increasing at a rapid rate~\cite{Black2007}. At the same time, there is a great opportunity to incorporate machine intelligence to manage these infrastructures efficiently. Yet, controlling the entire infrastructure in a centralized manner becomes untenable with the rise in the number of controllable entities in the system. 

Electric utilities, in particular, are at the cusp of a water-shed moment, where they will experience an enormous changes with the onset of an era when electric vehicles, grid level energy storage, and prosumers who have the resources to privately feed the grid via local generation (for instance, solar panels and storage)~\cite{parag2016electricity}. This will increase the dimension of the controllable space and upset the status quo. With the increase in the number of edge-network devices, the idea of allowing for an open energy market where each of the so-called \emph{prosumers} that may both produce and consume the commodity under question, has gained momentum. This notion is often referred to as Transactive Energy (TE)~\cite{Liu2017}. As such, several control frameworks for TE have been proposed. While originally TE was often tied to a new congestion pricing model for power utilities, these ideas have come increasingly under scrutiny. As grid operators, and utility in particular, can only operate behind the meter, ensuring the dynamic balance of supply and demand of power in real time as well as the security of such an open grid, remains an open question. This is mainly due to the vulnerability of the edge devices to external interventions, and the consequences of these vulnerabilities in a highly connected system are far reaching and pose damages beyond the meter~\cite{Mylrea2017}. 

\subsection{Background}
Recently, TE has gained traction again as an application for distributed Blockchain~\cite{Plaza2018,Mylrea2017,Munsing2017,Pipattanasomporn2018,Luo2019}, the same trust-less technology used for crypto-currencies. Blockchain platform to develop TE solutions come in two forms: permissionless/public or permissioned \cite{Cash2018}. Public blockchain platforms provide a level of trust in what are, arguably, otherwise completely trustless environments. Bitcoin, Ethereum, ripple, hydrachain and many other platforms are open to anyone who may want to participate, even anonymously. To prevent tampering with the content of the ledger, public Blockchains requires proof-of-work, proof-of-stake or proof-of-authority consensus algorithms so that these completely trustless environments can work. Permissioned Blockchains rely primarily on an authentication mechanism that manages the  authorizations for appending new transactions on the ledger.  Literature review shows the application of smart metering infrastructure in designing a Blockchain empowered TE system. Work done in \cite{Plaza2018} provides an architecture that connects IoT devices to smart meter's ports. The distribution system operator uses this \emph{metering} infrastructure as the role of the \textit{oracle} in the architecture. The authors in \cite{Pao2018} proposed a Blockchain based distributed ledger for storing the data acquired from metering devices as energy transactions in a secured and tamper proof manner. In \cite{Mylrea2017,Munsing2017} the authors proposed a Blockchain based meter that updates the Blockchain by creating a unique timestamp block for verification in a distributed ledger whenever a transaction takes place. Authors in \cite{Gao2018} proposed smart contracts protocol between the smart meter and the the sovereign blockchain network where, the smart contracts are triggered and executed based on the activity detected; malicious or appropriate. In \cite{Kounelis2017}, authors proposed the idea of a middle-ware controller that works as an interface between the smart meter and the grid and also acts as an input source for smart contracts. 

All functional aspects of TE enabled by Blockchain, from the bidding and pricing to the billing phase, can be orchestrated running \emph{Smart Contracts} \cite{Pee2019} over the Blockchain distributed ledger network infrastructure. There are several benefits of such an implementation: all  transactions are, once stored on the ledger, transparent to the participants who have an identical copy of the ledger. New transactions are  \textit{hash-chained} when appended to the ledger, an operation that makes them immutable.  This improves the resiliency of the electric grid against forms of cyber-attacks aimed at hampering the integrity and availability of the data.

The focus of this paper, though, is not security from external nefarious interventions, but rather from threats that exist within a TE infrastructure where selfish players have an intrinsic incentive to cheat about their use or their production of the commodities that they are trading. To the best of the authors' knowledge, much of the literature on Blockchain TE fails to account for the physical validation of the ground truth. It is true that transactions, once recorded on the ledger, are signed off transparently and the record is immutable. However, without any metering record that corroborates the physical delivery of the transaction amounts, the report in the ledger goes essentially unverified.  That is, Blockchain, in its current form, is not equipped to deal with such transgressions.

\subsection{Contribution}
In this paper, we assume that any participant in the TE marketplace might be a selfish or malicious entity trying to exploit vulnerabilities to gain an unfair advantage. 
To keep participants honest, we assume the stakeholder have access to sensor measurements that represent a distributed version of an advance metering infrastructure. These sensors measure various physical quantities from the grid and, as long as they are all consistent with the laws of physics, this is sufficient to verify that the power injections schedules produced during the TE market clearing process was honored. 
Since the sensing infrastructure is distributed, and selfish participants may want to misreport measurements, we take inspiration from the scheme proposed in \cite{vukovic2014security} in the context of power system State Estimation (SE) across different Regional Transmission Operators (RTO) and we adapt it to our verification task. The scheme includes in the SE the computation of a \textit{trust score} designed detect False Data Injection (FDI) attacks. Our underlying idea is to extend this approach to ensure that the proposed distributed verification process for a TE Blockchain can be carried out in a trust-less fashion. 

We note that FDI attacks in cyber-physical systems in general and on SE for power systems in particular, have attracted a tremendous amount of publications~\cite{liu2011false,liang20162015,ramakrishna2019detection}; a brief self-contained introduction is provided in Section~\ref{sec:Threat}.

To summarize, the main point of this paper is to address the verification of the transactions beyond the digital realms of the ledger in a trust-less and distributed environment, considering that selfish prosumers should be kept honest and that requires monitoring the physical operating state of the network. Towards this end, a scheme that should be incorporated between the market clearing phase and the billing phase is proposed and referred to as \textit{Robust State Verification}. 

The rest of the paper is organized as follows. In Section~\ref{sec:SysMod}, we describe the cyber-physical infrastructure. In Section~\ref{sec.physical-ver}, the laws of physics with regard to the power flow are defined. In Section~\ref{sec:verification}, the decentralized verification problem is presented and a corresponding Robust State Verification algorithm is proposed. In Section~\ref{sec:numericals}, numerical simulations are presented. Finally, we conclude in Section~\ref{sec:conclusion}.  

\textbf{Notations}: Boldfaced lower-case (resp.\ upper-case) letters denote vectors (resp.\ matrices) and  $x_i$ ($X_{ij}$) denotes the $i$th element of a vector $\bm{x}$ (the $ij$th entry of a matrix $\bm{X}$). Calligraphic letters are sets and $|\mathcal{A}|$ denotes the cardinality of a set $\mathcal{A}$.


\section{System Model}
\label{sec:SysMod}
\begin{figure}[htbp!]
    \centering
    \includegraphics[width=0.45\textwidth]{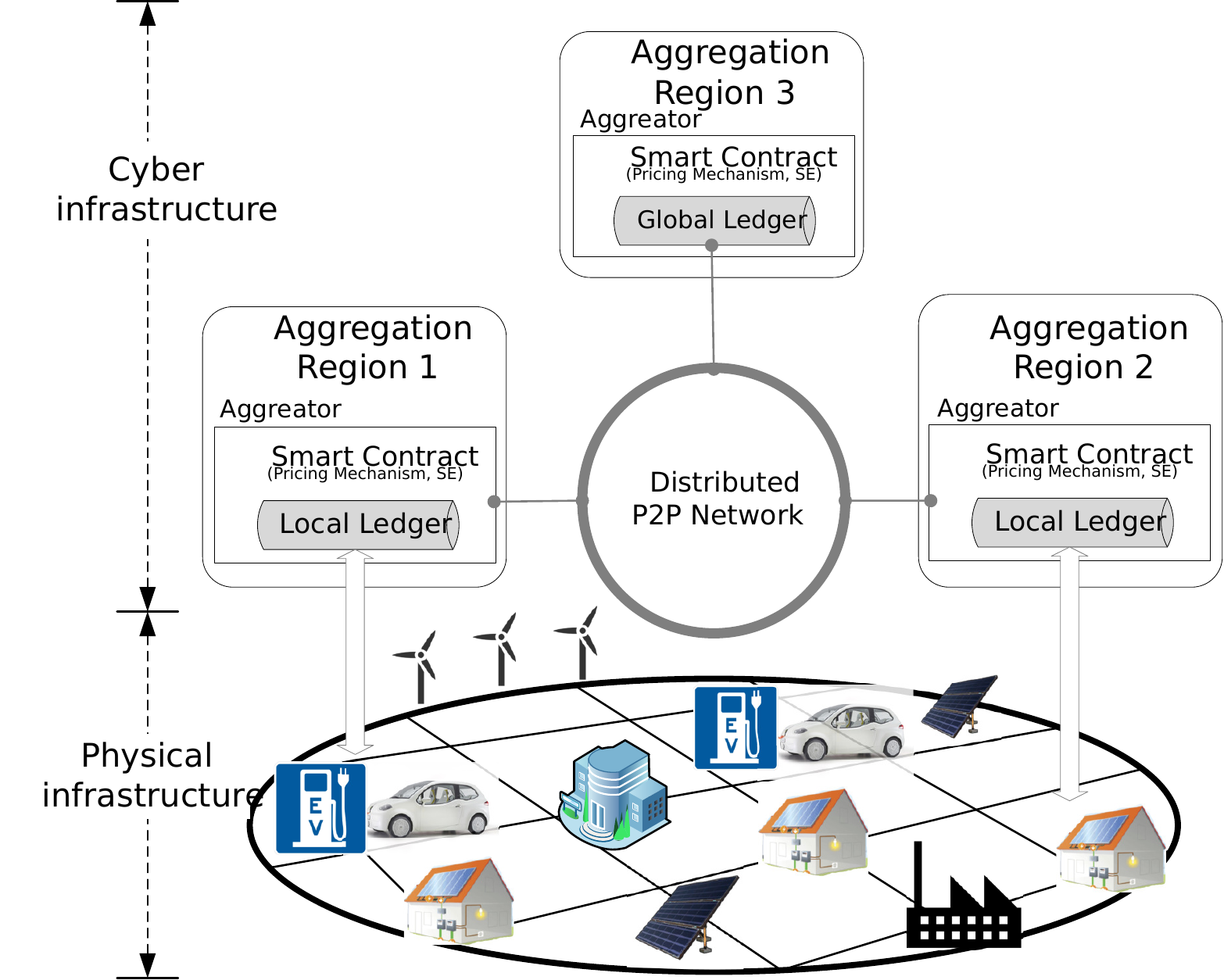}
    \caption{The Blockchain TE architecture}
    \label{fig:CPS}
\end{figure}

In this section, a brief description of the entire system is provided. The TE framework is split into two major layers. The lower layer comprising of the physical \textit{electric grid} with the various prosumers and participants distributed and connected throughout the grid, and the upper one is the cyber communication layer. Thus, TE is a Cyber-Physical System (CPS) as illustrated in Figure~\ref{fig:CPS}. The components are described in detail in the rest of this section.
\subsection{Physical System}
\begin{itemize}
  \item An electrical grid, which represents the physical infrastructure for the exchange of electricity. The network can be represented by a graph $\mathcal{G}_e = (\mathcal{B},\mathcal{E}_e)$, where $\mathcal{B}$ is the set of nodes (\textit{buses}) and $\mathcal{E}_e$ is the set of edges (\textit{lines}) connecting the network. Lines are characterized by admittance parameters $y_{ij},~\forall~ij\in \mathcal{E}_e$.
  \item Flexible loads (appliances) and distributed generation and storage assets, which connect to buses on the electrical network. \\
  \textbf{Note}: For the sake of the formulation, we assume there can only be one prosumer per bus $b$, knowing that we can denote multiple prosumers on a single bus by multiple nodes separated by zero impedance edges.
  \item Electrical sensors and control equipment.
\end{itemize}

\subsection{Cyber System}
The cyber system is comprised of an application layer which includes participants in the transactive market, secure policies for settling their transactions, and resources for computation, communication and data archival based on Blockchain. 
More specifically, through smart-contracts, a TE Blockchain implementation generally support and record three main collective decisions: 1) the \textbf{market clearing} decision process; 2) the \textbf{verification} of the transactions and 3) \textbf{billing}.

To improve scalability and maintain privacy through decentralization, the buses in the grid are spilt into a set $\mathcal{N}$ of $N$ \textit{aggregator} regions each of which serves a multitude of interconnected devices. $\mathcal{B}_n$ represents the set of buses $b$ belonging to an aggregator region $n$. Here, $\mathcal{B}_i \cap \mathcal{B}_j = \emptyset,~\forall~i,j \in \mathcal{N}$.

\begin{itemize}
    \item At the bottom sub-layer, the electric grid is divided into the several regions and includes \textit{prosumer} agents, which can buy and/or sell power and can control flexible loads, storage and generators that are connected to the bus subsets $\mathcal{B}_n$ above; this is the set of stakeholders in the TE market;
    \item The top sub-layer include \emph{aggregators} ${\cal N}=\{1,\ldots,N\}$ which operate on behalf of all prosumers that are connected to one sub-set of buses ${\cal B}_n, n\in {\cal N}$. There is one aggregator per region, and aggregators can coordinate the entire market because regions are interconnected and can communicate. 
    \item The coordination takes place over a communication graph $\mathcal{G}_c = (\mathcal{N},\mathcal{E}_c)$ where $\mathcal{E}_c$ are the communication links and an edge $ij \in \mathcal{E}_c$ implies that regions $i$ and $j \in \mathcal{N}$ share common state variables (this will be explained in detail in Section~\ref{sec.physical-ver}). The set of neighbors of a region $i \in \mathcal{N}$ is denoted by $\mathcal{N}_i$.
\end{itemize}
The assumption that the system is partitioned in two layers is without loss of generality, as one can always restrict the agents to a single home-feeder flattening the architecture and still fit in the model. Of course the finer is the sectioning the slower are the peer-to-peer network processes. 

Communication of information among the prosumers via the buses to the aggregators and between the aggregators are executed via private writing and reading on the distributed ledgers of the Blockchain, which also contain as smart-contracts the algorithms and policies that govern the TE market: 
\begin{itemize}
    \item Prosumers in each region can access the local (regional) ledger. 
    \item Aggregators can access both the local (regional) ledger as well as the global (common) ledger. 
\end{itemize}
Through the abstraction of a distributed ledger, communication, computation, and data archival components are unified on the Blockchain, providing all the inherent benefits of security as expected on a Blockchain.


\section{Electric Grid Constraints}\label{sec.physical-ver}
Before presenting the algorithm for achieving consensus on what the power injections from the various prosumers in actuality were, here we introduce the physical behavior and governing equations of our electric grid model that will be used as a tool for verification.   

\subsection{Physical constraints}\label{sec.physical}
Physical constraints relate the sensor measurements through the laws of physics of the electrical grid. Any measurable physical quantity in the system can be derived from the voltage phasors (amplitude and phase of the AC voltage at each bus) and the grid parameters (admittances). The voltage phasors are, in fact, called the \textit{state} of the system, and may be measured directly by employing expensive Phasor Measurement Units (PMUs). In the absence of PMUs, however, one can measure a number of variables. Since we focus on distribution systems and the goal we pursue is not state estimation but rather verifying the injections are close to what the market established as the schedule, here we use the DistFlow equations \cite{baran1989optimal,barenwucapplacement}, that include quantities more easily metered.
Let an edge in the electrical grid be denoted by $\ell =(b,b')\in \mathcal{E}_e$, where $b,b'\in \mathcal{B}$. Defining the \textit{from} and \textit{to} functions which return the source node and incident node for an edge respectively, such as:
\begin{equation*}
fr(\ell)=b,~~to(\ell)=b'    
\end{equation*}
The inverse function $to^{-1}(b)$ gives back the edge pointing to bus $b$ (for a radial graph the \textit{to} bus $b$ for each edge is unique) and $fr^{-1}(b)$ returns all the edges originating at bus $b$. 

Within a time horizon $\mathcal{T}$, at a time $t \in \mathcal{T}$, the power flow equations at each of the branches $\ell \in \mathcal{E}_e$ are given by:
\begin{subequations}\label{eq:dist_flow}
\begin{align}
    p_{to(\ell)}(t)&=P_{\ell}(t)-\!\!\!\!\!\!\!\sum_{\ell'\in fr^{-1}(to(\ell))}\!\!\!\! P_{\ell'}(t)-\Re(z_{\ell})c^2_{\ell}(t)
    \label{eq:PF1}\\
    q_{to(\ell)}(t)&=Q_{\ell}(t)-\!\!\!\!\!\!\!\sum_{\ell'\in fr^{-1}(to(\ell))}\!\!\!\! Q_{\ell'}(t)-\Im(z_{\ell})c^2_{\ell}(t)\label{eq:PF2}\\
    v_{fr(\ell)}^2(t)&=v_{to(\ell)}^2(t)+
    2\left(\Re(z_{\ell})P_{\ell}(t)+\Im(z_{\ell})Q_{\ell}(t)\right)\nonumber\\&~~-|z_{\ell}|^2c_{\ell}^2(t)\label{eq:PF3}\\
    v^2_{fr(\ell)}(t)&=\frac{P^2_{\ell}(t)+Q^2_{\ell}(t)}{c^{2}_{\ell}(t)}.\label{eq:PF4}
\end{align}
\end{subequations}
All these equations are linear in the bus and branch quantities $(p_b(t),q_b(t),v_b^2(t))$ (respectively the real and reactive power injections, and the squared voltage magnitudes) and $(P_{\ell}(t),Q_{\ell}(t),c^2_{\ell}(t)
)$ (respectively the real and reactive power flows, and the squared current magnitudes) except equation \eqref{eq:PF4}. However, including an auxiliary variable 
$x'_{\ell}=(P^2_{\ell}+Q^2_{\ell})/{c^{2}_{\ell}}$ can simplify the description of the physical constraints. 
Let $\bm{x}(t)$ be the vector of system variables at time $t$ which contains blocks of the injection variables and branch variables. For the particular set of equations used in this paper shown in equation~\eqref{eq:dist_flow}, the injection variables are $(p_b(t),q_b(t), v_b^2(t))^{\text{T}},~\forall b \in \mathcal{B}$, and the branch flow variable are $(P_l(t), Q_l(t), c_l^2(t))^{\text{T}},~\forall ~l \in \mathcal{E}_e$.
Let $\bm{x}(t)$ also include the auxiliary variables and, partition this vector in two parts: $\bm{x}_a(t)$  containing those variables whose measurements are available and and $\bm{x}_u(t)$ including the variables that are not measured. The measurements are noisy versions of the variables in $\bm{x}_a(t)$, i.e.: 
 \begin{equation}\label{eq.sensors}
 \bm{s}(t) = \bm{x}_a(t) + \bm{w}(t), 
 \end{equation}
which means that, within a margin of error that depends on the noise, they should meet the physical constraints of the grid given in equation set~\eqref{eq:dist_flow}. 
These physical constraints are a set of non-linear homogeneous equations; we refer to them in vector form as follows:
\begin{equation}\label{eq:elec_grid_const}
    \bm{h}(\bm{x}(t)) = \bm{h}(\bm{x}_a(t),\bm{x}_u(t)) = \bm{0},~~~\forall t \in \mathcal{T},
\end{equation}
where the definition of $\bm{h}(\cdot)$ is some appropriate vector function.
Relaxing the constraints by ignoring the non-linear relationships among the auxiliary variables and the remaining entries of the vector $\bm{x}$, and without loss of generality equations \eqref{eq:PF1}$-$\eqref{eq:PF4} can be written in the linear form:
\begin{equation}\label{eq:linear-constr}
    \bm{H}\bm{x}=\bm{H}_a\bm{x}_a+\bm{H}_u\bm{x}_u=\bm{0}, 
\end{equation}
The physics of the measurements is represented by  $\bm{s}=\bm{x}_a+\bm{\epsilon}$ and the physics of the system implies that $\bm{x}_a$ satisfies \eqref{eq:linear-constr}. 

Note that the system variables can be further partitioned per aggregator region. Let $\bm{x}^{(i)}(t)$ be the copies of variables in $\bm{x}(t)$ pertaining to aggregator region $i$. Since there are tie-lines among the regions some of the variables in $\bm{x}^{(i)}(t)$ will have identical counterparts in the $\bm{x}^{(j)}(t)$ of a neighboring region $j$. Measurements from an area $\bm{s}^{(i)}$ cannot be independently verified unless one has branch measurements available; even then, an aggregator should not trust other aggregators. If $m_i$ is the number of variables pertaining to aggregator region $i$, then $\bm{D}^{(i)} \in \mathbb{R}^{m_i \times m_i}$ is a diagonal matrix with $[\bm{D}^{(i)}]_{kk}$ equal to the number of regions with which region $i$ has the $k$-th variable of $\bm{x}^{(i)}$ in common with, $\forall k \in [m_i]$. Similarly $\overline{\bm{D}}^{(i)}\in \mathbb{R}^{m_i \times m_i}$ is also a diagonal matrix such that $[\overline{\bm{D}}^{(i)}]_{kk} =1/[\bm{D}^{(i)}]_{kk}$ if $[\bm{D}^{(i)}]_{kk}\neq 0$ and $[\overline{\bm{D}}^{(i)}]_{kk} =0$ if $[\bm{D}^{(i)}]_{kk} = 0$.

From the vantage point of each region $i \in \mathcal{N}$, only a subset of the variables $\bm{x}_a^{(i)}$ are measured, so that $\bm{s}^{(i)}=\bm{x}_a^{(i)}+\bm{\epsilon}^{(i)}$. 
Also, not all constraints in  \eqref{eq:linear-constr} include the variables $\bm{x}^{(i)}$. By isolating the equations that involve buses/lines in region $i$ the following can be written:
\begin{equation}
    \label{eq:linear-constr_region}
    \bm{h}^{(i)}(\bm{x}^{(i)})=\bm{H}^{(i)}\bm{x}^{(i)}=\bm{0}.   
\end{equation}
Equations~\eqref{eq:linear-constr_region} along with sensor measurements $\bm{s}(t)$ in \eqref{eq.sensors} play a key role in verifying whether the prosumers' power injections match the scheduled transactions, allowing to correctly bill each prosumer for their actual usage, and possibly fine prosumers for schedule violations. 

Next, in Section~\ref{sec:verification}, we provide a decentralized algorithm that allows aggregators to use equations~\eqref{eq:linear-constr_region}, along with their local portion of sensor measurements $\bm{s}^{(i)}$ to cross verify the actual power injections in the system. 


\section{Robust Verification Algorithm}\label{sec:verification}
A distributed market breeds an inherent selfishness among the various players involved, i.e, the prosumers might want to pay less for what they consume or charge more for what they produce, and alternatively, they might wish to cheat the system by not producing what they promised during the pricing stage. Hence, once the market is cleared, we have the power injections scheduled as $\mathfrak{\bm{p}}^\star(t)$, measurements are needed to verify the actual power injections ${\bm p}_b(t)$. But because no-one can be absolutely trusted to self-police, it is necessary to stitch measurements together through the power flow constraints in Section \ref{sec.physical-ver}.  
\begin{figure}[!htbp]
\centering
    \includegraphics[width=0.9\linewidth]{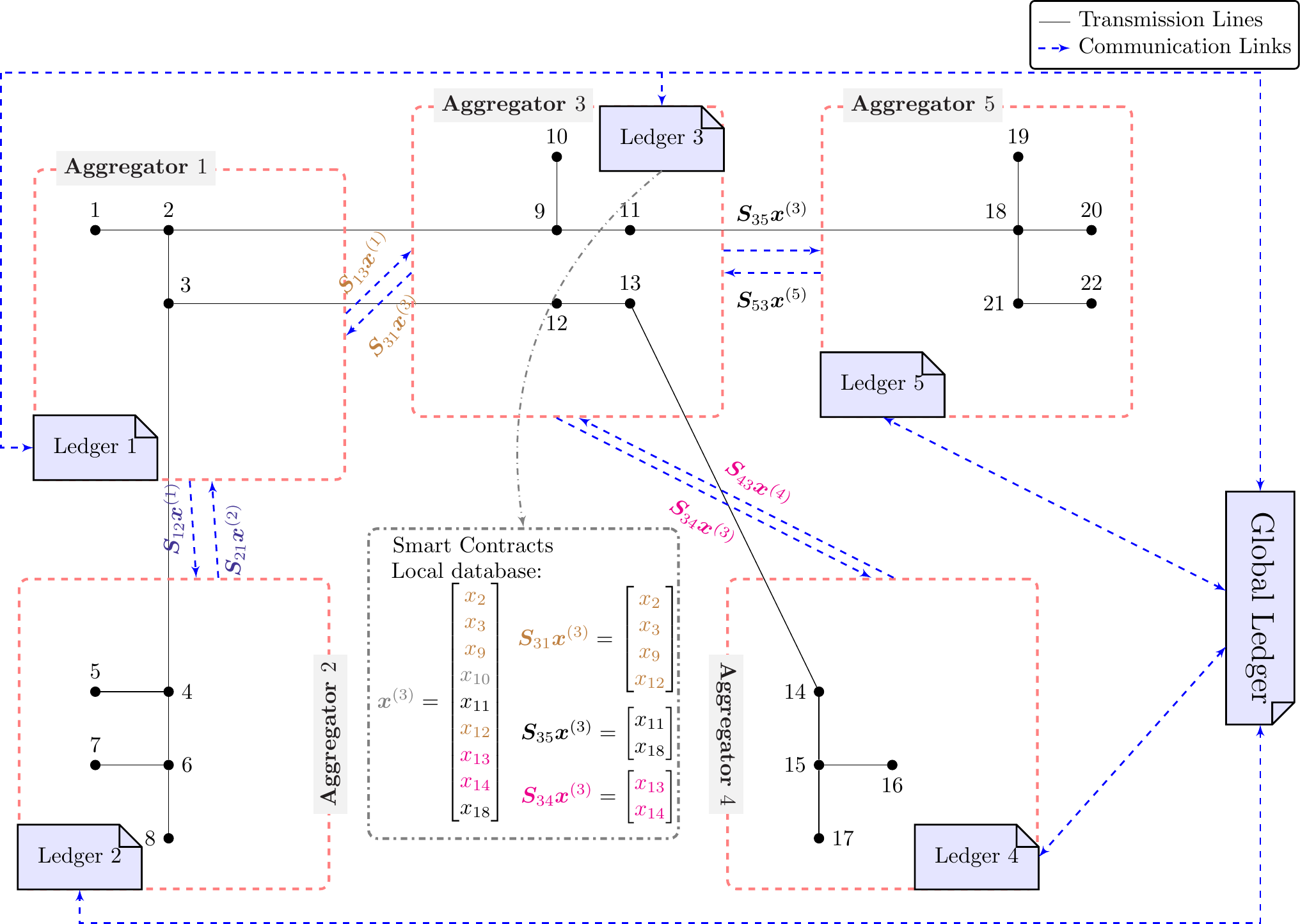}
\caption{State Verification Architecture for a test case with five aggregator regions.}
\label{fig:state_est}
\end{figure}
The problem of verification is similar to the state estimation problem in power systems monitoring~\cite{liu2011false}, which concerns the estimation of all the state variables (voltage phasors) from limited measurements. The goal, however, is fundamentally different: it is to cross-validate self-reported power injections, to ensure the billing reflects what was actually exchanged and penalizes discrepancies that jeopardize the physical system stability.  As mentioned in Section~\ref{sec:intro}, the problem of FDIAs in state estimation, centralized or distributed, has been widely studied. In this paper, for the verification algorithm, a fully decentralized implementation of Alternative Direction Method of Multipliers (ADMM)~\cite{boyd2011distributed,kekatos2012distributed} is employed since the grid system is partitioned into regions whose goal is to individually estimate the state variables within the region and agreeing about the variables related to the branches that tie different regions. An illustration of the system is shown in Figure \ref{fig:state_est}. It is important to remark that we need to have redundancy in the measurements to ensure Byzantine Fault Tolerance for the verification process. 

\subsection{Problem Formulation}
Let $\bm{S}^{(i)}$ be the selection matrix that selects the system variables in $\bm{x}(t)$ that pertain to the $i$th region, such that $\bm{x}^{(i)}(t) = \bm{S}^{(i)}\bm{x}(t)$. Furthermore, let  $\bm{S}^{(i)}_a$ be the selection matrix that selects from $\bm{x}(t)$ those variables pertaining to region $i$ whose measurements are available, i.e., $\bm{x}^{(i)}_a(t) = \bm{S}^{(i)}_a\bm{x}(t)$, and similarly let $\bm{S}_p^{(i)}$ be a selection matrix that extracts the active power variables in the vector $\bm{x}^{(i)}(t)$. The \textit{state verification problem} can be formulated as the following optimization, written in the form amenable to the ADMM decomposition:
    \begin{align}
        \min_{\{\bm{x}^{(i)}(t)\}^{t\in\mathcal{T}}_{i\in \mathcal{N}}}~&\!\!\!
        \sum_{t\in\mathcal{T}}
        \sum_{i\in\mathcal{N}}
        \left(\|\bm{s}^{(i)}(t) - \bm{x}^{(i)}_a(t) \|^2
        +\right. \label{eq:State_Est}\\
           &\!\!\!+\|\bm{h}^{(i)}(\bm{x}^{(i)}(t) )\|^2_2  + c_1\left.\|{\bm{\mathfrak{p}}^\star}^{(i)}(t)  - \bm{S}^{(i)}_p\bm{x}^{(i)}\|^2_2\right)\nonumber\\
        \text{s.t.} ~&~ \bm{S}_{ij}\bm{x}^{(i)}(t) = \bm{S}_{ji}\bm{x}^{(j)}(t), \forall ij \in \mathcal{E}_p,\label{eq.consensus}\\
        ~&~ \bm{x}^{(i)}(t)=\bm{S}^{(i)}\bm{x}(t)~,~~\forall i\in\mathcal{N}\\
        ~&~
        \bm{x}_a^{(i)}(t)=\bm{S}_a^{(i)}\bm{x}(t), ~~\forall i\in\mathcal{N}.
    \end{align}
where $[{\bm{\mathfrak{p}}^\star}^{(i)}(t)]_k = p_{b_k}(t),~\forall b_k \in \mathcal{B}_i$ is the scheduled power flow injections at time $t$ for all bus in region $i$, and the consensus among the common variables across neighboring regions is enforced through \eqref{eq.consensus}; more specifically, $\bm{S}_{ij}$ is the selection matrix that extracts from $\bm{x}^{(i)}$ the common variables between neighboring regions $i$ and $j$, and the definitions of the functions $\bm{h}^{(i)}(\cdot)$ are given in Section~\ref{sec.physical-ver}. In words, the objective is to find the the values of the system variables that have the minimum residual error with respect to the measurements, the scheduled injection and fit the physical model equations with the least residual error, measured in terms of squared norm. The solution is needed for the verification since the components of the solution $\bm{S}^{(i)}_p\bm{x}^{(i)}$ that correspond to the bus power injections are used in the Blockchain TE as the closest approximation for the ground truth for billing; also the severity of their deviations  from the schedule $\mathfrak{p}^\star_b(t)-p_b^a(t)$ determines the penalties assigned to the prosumers using the specific bus.  

An iterative process to arrive at the minimizer of the problem in~\eqref{eq:State_Est} is as follows: for $\mathcal{T} = \{0\}$
\begin{align}
    \bm{x}^{(i)}_{k+1}(t) =& \left( {\bm{H}^{(i)}}^\text{T}\!\bm{H}^{(i)} \!+\! {\bm{S}^{(i)}_a}^\text{T}\bm{S}^{(i)}_a \!+\! {\bm{S}^{(i)}_p}^\text{T}\bm{S}^{(i)}_p \!+\! c_2\bm{D}^{(i)}\right)^{-1}\nonumber\\
    \times~&\!\!\!\left( {\bm{S}^{(i)}_a}^\text{T}\bm{s}^{(i)}(t) + {\bm{S}^{(i)}_p}^\text{T}{\bm{\mathfrak{p}}^\star}^{(i)}(t) + c_2\bm{D}^{(i)}\bm{\upsilon}^{(i)}_k(t) \right)\label{eq:ADMM_x},\\
    \bm{\psi}^{(i)}_{k+1}(t) =& \overline{\bm{D}}^{(i)} \sum_{j: ij \in \mathcal{E}_p} \bm{S}_{ij}^\text{T}\bm{S}_{ji}\bm{x}^{(j)}_{k+1}(t),\label{eq:ADMM_phi}\\
    \bm{\upsilon}^{(i)}_{k+1}(t) =& \bm{\upsilon}^{(i)}_{k}(t) + \bm{\psi}^{(i)}_{k+1}(t) - 0.5\left(\bm{\psi}^{(i)}_{k}(t) + \bm{x}^{(i)}_{k}(t) \right)\label{eq:ADMM_ups},
\end{align}
for all $k\geq 0$ until a termination condition is satisfied, while the ADMM states $\bm{x}^{(i)}_0(t),\bm{\psi}^{(i)}_0(t),$ and $\bm{\upsilon}^{(i)}_0(t)$ initialized arbitrarily, to $\left(\overline{\bm{D}}^{(i)} \sum_{j: ij \in \mathcal{E}_p} \bm{S}_{ij}^\text{T}\bm{S}_{ji}\bm{x}^{(j)}_{0}(t)\right)$, and to $\frac{1}{2}\left(\bm{\psi}^{(i)}_{0} + \bm{x}^{(i)}_{0}(t) \right)$ respectively. An algorithmic implementation of the iterative process is presented in Algorithm~\ref{alg:state_estimation} [ignore Steps 3, 4, 6, 19, and 21 for now]. Figure~\ref{fig:state_est} shows a pictorial representation of the mechanisms of the algorithm. Buses in each aggregator region $i$ have access to their local ledgers $LL_i$. The aggregators, in addition to their local ledger, also have access to the Global Ledger $GL$. The aggregator collects measurements from their buses and stores them on $LL_i$. Iterates after each ADMM update at an aggregator are also saved in $LL_i$.
Aggregators exchange parts of their states that correspond to variables on the tie lines with a neighbor. 

\begin{algorithm}
\caption{Robust State Verification; Here $LL_i~\forall i \in \mathcal{N}$ and $GL$ represents the local and global ledgers respectively. The symbols $\Leftarrow$ and $\Rightarrow$ correspond to upload and download to and from a ledger, respectively.}
\label{alg:state_estimation}
\begin{algorithmic}[1]
\State $LL_i \Leftarrow$ \text{ Collect local measurements} $\bm{s}^{(i)}(t)$
\State Initialize the ADMM states according to~\cite{vukovic2014security}.
\State Initialize trust score $\bm{\pi}= \bm{0}$
\State Initialize disagreements $d_{ij} = 0,~\forall ij \in  \mathcal{E}_p$
\Repeat
    \State $\bm{\pi}^{-} \gets \bm{\pi} \Leftarrow GL$
    \State ADMM States $\Leftarrow LL_i$
    \State $[\bm{x}^{(i)}(t)]^{-}\leftarrow \bm{x}^{(i)}(t)$
    \State ADMM update of $\bm{x}^{(i)}(t)$
    \State $LL_i \Leftarrow \bm{x}^{(i)}(t)$ 
        \ForAll{$j: ij\in \mathcal{E}_p$}
            \State Open private channel $GL_{ij}$
            \State $GL_{ij} \Leftarrow \bm{S}_{ij}\bm{x}^{(i)}(t)$
        \EndFor
        \ForAll{$j: ij\in \mathcal{E}_p$}
            \State  $ \bm{S}_{ji}\bm{x}^{(j)}(t) \Leftarrow GL_{ji}$
        \EndFor
        
        \State $LL_i \Leftarrow$ Update intermediate states of the ADMM.
        \State Run Algorithm~\ref{alg:Security} to update $\bm{\pi}$ and $d_{ij},~\forall ij \in  \mathcal{E}_p$.\label{alg:step:alg2}
\Until{$\|\bm{\pi}\! -\! \bm{\pi}^{-}\|_\infty \!\! \leq \!\! \epsilon_\pi$ or $\|\bm{x}^{(i)}(t)\! -\! [\bm{x}^{(i)}(t)]^{-} \|_\infty\! \leq\! \epsilon$}
\State Restart the algorithm with $\{\mathcal{G}_p\setminus  \left(\arg\max_i \bm{\pi}\right)\}$\label{alg:step:restart}
\end{algorithmic}
\end{algorithm}

\subsection{Threat Model and Robust State Verification}\label{sec:Threat}

We assume that the adversary (a threat agent or a group of coordinating agents) is an insider who has legitimate physical and logical access to the network and ledgers through the certification mechanism. We also assume that the adversary has the capability to manipulate sensors measurements, either by compromising the sensors or the communication between the sensors and the aggregators. The motivation of attackers is trying to cheat the systems in order to obtain the financial benefits or disrupt the verification process.

In fact, sensor values can be forged or spoofed.
If a malicious user is in region $j$, it may modify the values of the input measurements, altering aggregator's $j$ measurement vector by perturbing the measurement vector as follows:
\[
    \widetilde{\bm{s}}^{(j)} = \bm{s}^{(j)} + \bm{a}^{(j)},
\]
where the perturbation $\bm{a}^{(j)}$ has non zero entries in the locations that correspond to the false sensor measurements. Also, if the aggregator itself acts maliciously, it can inject false data into the updates of $\bm{S}_{ji}\bm{x}^{(j)}$ that are passed to a neighbor $i$ (see equation~\eqref{eq:ADMM_phi}). This leads to the discrepancies in the updates of the neighboring aggregators,
\[
    \begin{split}
        \widetilde{\bm{x}}^{(i)}_{k+1}(t) =& {\bm{x}}^{(i)}_{k+1}(t) + c_2\bm{M}\bm{D}^{(i)}\overline{\bm{D}}^{(i)}\bm{S}_{ij}^\text{T}\bm{S}_{ji}\bm{a}^{(j)}_{k}(t),
    \end{split}
\]
where 
\[\bm{M} = \left( {\bm{H}^{(i)}}^\text{T}\bm{H}^{(i)} + {\bm{S}^{(i)}_a}^\text{T}\bm{S}^{(i)}_a + {\bm{S}^{(i)}_p}^\text{T}\bm{S}^{(i)}_p + c_2\bm{D}^{(i)}\right)^{-1}.\]

If the vanilla ADMM updates as seen in equations~\eqref{eq:ADMM_x}--\eqref{eq:ADMM_ups} is used as is to solve \eqref{eq:State_Est}, the FDIAs are successful leading to divergence of the algorithm or in the worst case convergence to a false optimum. The former is a special case of \textit{Denial of Service} attack, where the aggregators are unable to complete the verification process.
When the algorithm converges to a false optimum, this leads to an FDIA to the ledger.

\begin{algorithm}
\caption{Detection Loop ;\\
$F(d_{ij}, \bm{S}_{ij}\bm{x}^{(i)}(t),$ $\bm{S}_{ji}\bm{x}^{(j)},~\forall j: ij\in \mathcal{E}_p$)}
\label{alg:Security}
\begin{algorithmic}[1]
\State Set $\alpha$
\State $GL \Leftarrow$ Update $d_{ij} \forall j: ij \in \mathcal{E}_p$ where,
\begin{multline*}
    d_{ij}  \leftarrow \frac{\alpha_k/4}{|\bm{S}_{ij}\bm{x}^{(i)}(t)| |\mathcal{T}|} \sum_{t\in\mathcal{T}}{\|\bm{S}_{ij}\bm{x}^{(i)}(t)\!-\!\bm{S}_{ji}\bm{x}^{(j)}(t)\|_F^2} \\+ (1-\alpha_k)d_{ij}
\end{multline*}
\State $GL \Rightarrow$ $\bm{B}$, where $[\bm{B}]_{ij} = \frac{d_{ij}}{\sum_{k: k \in \mathcal{E}_p} d_{ik}+\epsilon}$
\State Compute $\bm{\pi}$, the left eigenvector of $\bm{B}$, s.t. $\bm{\pi}^\text{T}\bm{B} = \bm{\pi}^\text{T}$.
\State Output $\bm{\pi}, d_{ij},~\forall ij \in \mathcal{E}_p$
\end{algorithmic}
\end{algorithm}

In order to detect the attack, we leverage the method proposed in \cite{vukovic2014security}. The methodology is presented in Algorithm~\ref{alg:Security}, which is a detection subroutine in Algorithm~\ref{alg:state_estimation} [Step~\ref{alg:step:alg2}] with initialization given in [Steps~3, 4].

In Algorithm~\ref{alg:Security}, each region $i$ calculates a measure of disagreement, $d_{ij}$, with a neighboring region $j$, and $\bm{B}$ is the matrix of normalized disagreement scores. A left eigenvector ($\bm{\pi}$) of $\bm{B}$ (equivalently, the stationary distribution of a Markov chain with probability transition matrix $\bm{B}$) can be calculated. In general, the presence of malicious agents will not lead to a convergence, and the algorithm will assign higher scores for those regions who are in maximal disagreement.
Being inspired by the work in \cite{vukovic2014security} that used the norm and location of the highest value of $\bm{\pi}$ as an indicator of the presence of an attack and the indices of the most likely attackers; to mitigate the impact of FDIA, step~\ref{alg:step:restart} is added to Algorithm~\ref{alg:state_estimation}, to restart the algorithm by isolating the identified attacker. It must be noted, however, that the method is more or less prone to miss identification errors depending on the connectivity of the electric grid graph.

\textbf{Termination conditions}: Algorithm~\ref{alg:state_estimation} terminates under two scenarios. In the first scenario, the algorithm terminates when the aggregator is in consensus with its neighboring aggregators and has converged to a point, thus satisfying $\|\bm{x}^{(i)}_k(t) - \bm{x}^{(i)}_{k-1}(t)\|_\infty \leq \epsilon$. The second scenario is met when all the aggregators have come to a consensus about the presence as well as the identity of an FDI attacker amongst them. We assume that in each iteration $k$, an individual aggregator calculates an \textit{excluded average} and  an \textit{excluded standard deviation} for all the aggregator $i\in\mathcal{N}$ using \eqref{eq:avg} and \eqref{eq:sd} respectively:
\begin{subequations}
\begin{align}
    \label{eq:avg}
    {\mu}_i(\bm{\pi}) &= \frac{1}{|\mathcal{N}|-1}\textstyle{\sum_{j\in\mathcal{N}\setminus\{i\}}} \pi_i \\
    \label{eq:sd}
    \sigma_i(\bm{\pi}) &= \sqrt{\frac{1}{|\mathcal{N}|-1}\sum_{j\in\mathcal{N}\setminus\{i\}} |\pi_i-\mu_i(\bm{\pi})|^2}.
\end{align}
\end{subequations}
Since all the aggregators have the same $\bm{\pi}$ vector, they all have the same mean and standard deviations regarding the other aggregators.
For the second scenario to trigger, the $\bm{\pi}$ vector has to converge in the norm-sense, i.e, the condition $\|\bm{\pi}_k - \bm{\pi}_{k-1}\|_\infty \leq \epsilon_\pi$ is satisfied, and $\pi_i > \mu_i(\bm{\pi}) + \beta_i\sigma_i(\bm{\pi})$ has to be true for some $\beta_i>0$ and $i\in \mathcal{N}$.

\textbf{Note on Stealth Attacks}: Furthermore, a stealth attack can be staged in some cases leading to a missed detection. The global measurement vector as a result of the attack is given by $\widetilde{\bm{s}} = \bm{s} + \bm{a}$. A stealth attack can be posed by choosing a sparse attack vector, $\bm{a}$, where the non-zero entries correspond to the sensors that are attacked, such that the constraint $\bm{h}(\bm{x}+\bm{a}) =0$ is satisfied even with the perturbed state, where $\bm{x}$ corresponds to the true variables.
Here, without any change in the loss function in problem~\eqref{eq:State_Est}, the attacker is still able to alter the output of the algorithm. Such attacks are extremely hard to detect in the absence of specially imposed structure on the true measurement vectors, let alone mitigate. These types of attacks are only possible when a malicious aggregator is able to gain complete knowledge about their neighbors' parameters. This constitutes the worst case scenario for an attack. Note that such attacks are not always possible in practice, depending on the sparsity pattern of the number of attacked sensors.

For example, consider the linear form of the power flow equation pertaining to a region $j$ as given in equation~\eqref{eq:linear-constr_region}. Suppose region $j$ is attacked, say via FDI to its measurements or to its ADMM update states $\bm{x}^{(j)}(t)$, and as a consequence of this attack, let $\widetilde{\bm{x}}^{(j)}(t) = \bm{x}^{(j)}(t) + \bm{a}^{(j)}$ be the modified output of its ADMM state update. For region $j$ to pose a stealth attack on, say, region $i$, the attack vector $\bm{a}^{(j)}$ has to be chosen carefully such that the elements of the vector are non-zero in the positions corresponding to the shared variables with region $i$. 
Furthermore, for this digression in the $\bm{x}^{(j)}(t)$ iterate to affect region $i$'s update of $\bm{x}^{(i)}(t)$ adversely (that is $\widetilde{\bm{x}}^{(i)}(t) = \bm{x}^{(i)}(t) + \bm{a}^{(j)}$), then $\bm{a}^{(j)}$ has to be in the null space of the system matrix $\bm{H}^{(i)}$. Even telling is the fact that $\bm{S}_{ij}\bm{a}^{(j)}$ has to be in the null space of $\bm{H}^{(i)}\bm{S}_{ij}^\text{T}$. To perform such an elaborate attack, region $j$ requires information regarding $\bm{H}^{(i)}\bm{S}_{ij}^\text{T}$ and moreover, there might not exist a non-trivial null space in $\bm{H}^{(i)}\bm{S}_{ij}^\text{T}$.
\begin{figure*}[htbp]
    \centering
    \includegraphics[width=0.78\textwidth]{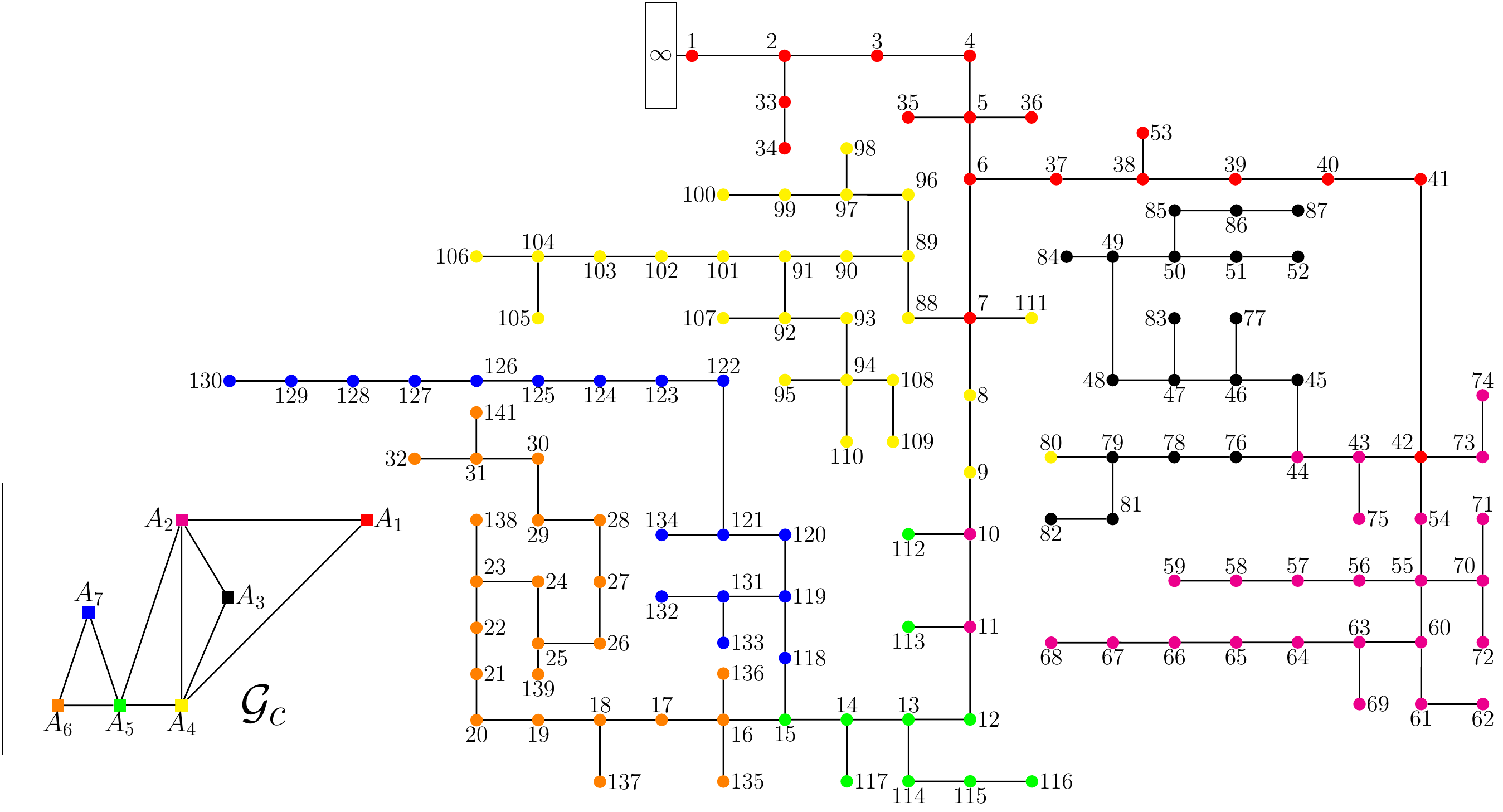}
    \caption{$\mathcal{G}_e$ is the network graph corresponding to the $141$ bus radial distribution feeder case. The network in the box in the bottom right shows the communication graph $\mathcal{G}_c$ with the nodes representing aggregators: $A_i\equiv$ Aggregator $i,~\forall i \in \mathcal{N}$.}
    \label{fig:case141}
\end{figure*}


\section{Simulation Results}
\label{sec:numericals}
In this section, the workings of the methodology proposed in Section~\ref{sec:verification} is illustrated using a $141$ bus radial distribution feeder case \cite{Matpower}. The case consists of $141$ buses with bus $1$ considered to be a substation with infinite capacity. The electric grid is divided into $N = 7$ aggregator regions, and the set of buses belonging to different regions are indicated using colors in Figure~\ref{fig:case141}. The communication graph $\mathcal{G}_c$ is also shown in Figure~\ref{fig:case141}; the structure is a consequence of the assignment of buses into different aggregator regions.
\begin{figure}[htbp!]
	\centering
	\subfloat[]{%
	\label{fig:noattack}%
	\includegraphics[width=0.48\linewidth]{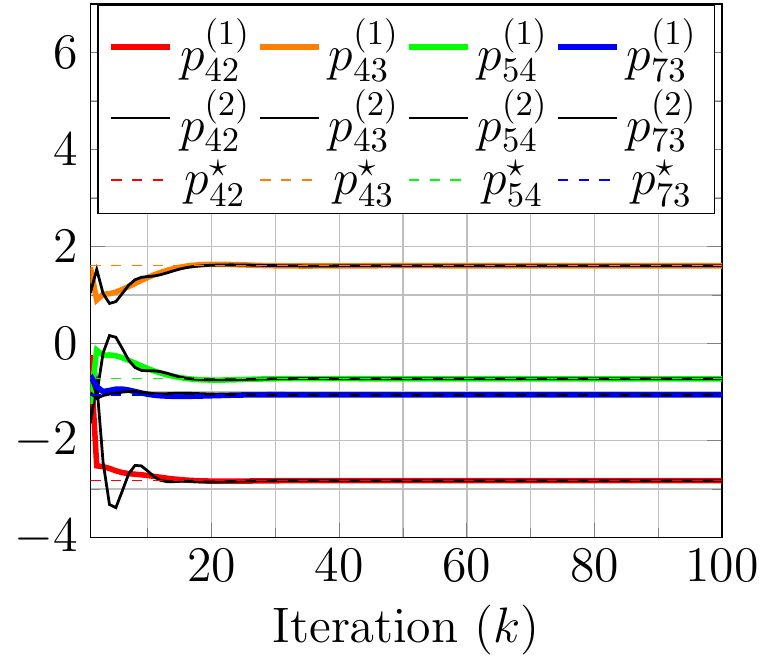}%
	}%
	\quad
	\subfloat[]{%
	\label{fig:attack}%
	\includegraphics[width=0.45\linewidth]{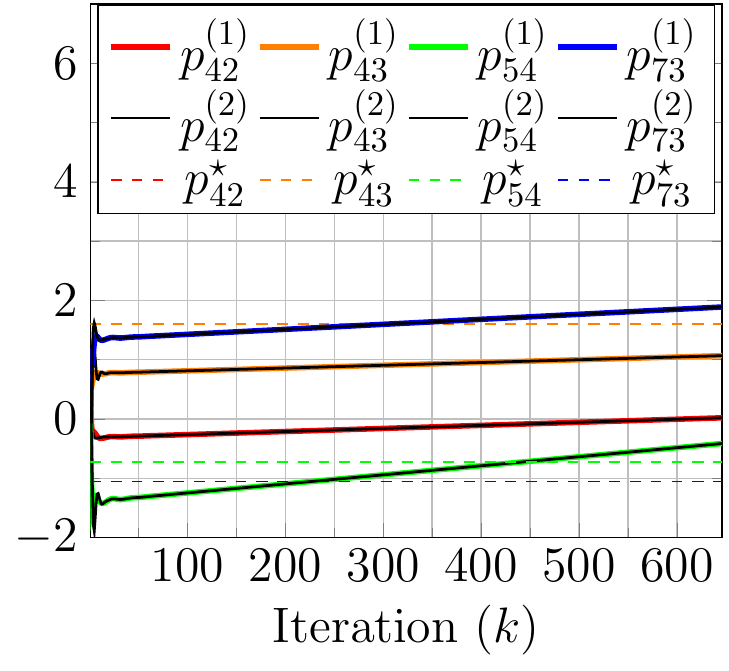}%
	}%
	\caption{\protect\subref{fig:noattack} Consensus and convergence of real power injections (in MW) by common variables of aggregators $1$ and $2$ when there is no attack. \protect\subref{fig:attack} Consensus but divergence of real power injections (in MW) by common variables of aggregators $1$ and $2$ when aggregator $1$ is an attacker. In this case, the verification process can not be completed. The common variables correspond to buses $42,43,54,$ and $73$ in Figure~\ref{fig:case141}. Aggregator $i$'s estimate of the real power injection at bus $b$ is denoted by $p_b^{(i)}$.}
	\label{fig:numericals}
\end{figure}
\par For the simulations, the parameters were chosen as follows: $\epsilon_\pi = 1e-3$, $\epsilon=1e-3$, $\alpha_k = 1/k$, $c_1=c_2=0.5$, $\beta_i=2,~\forall i \in \mathcal{N}$. The measurements of the available variables, $\bm{x}_a$, were noisy versions of MATPOWER output of the power flow calculations for the $141$ bus radial distribution feeder case. The noise, $w_i$ was chosen to be i.i.d Gaussian with zero mean and variance $1$.
\begin{figure}[!htbp]
    \centering
    \includegraphics[width=0.98\linewidth]{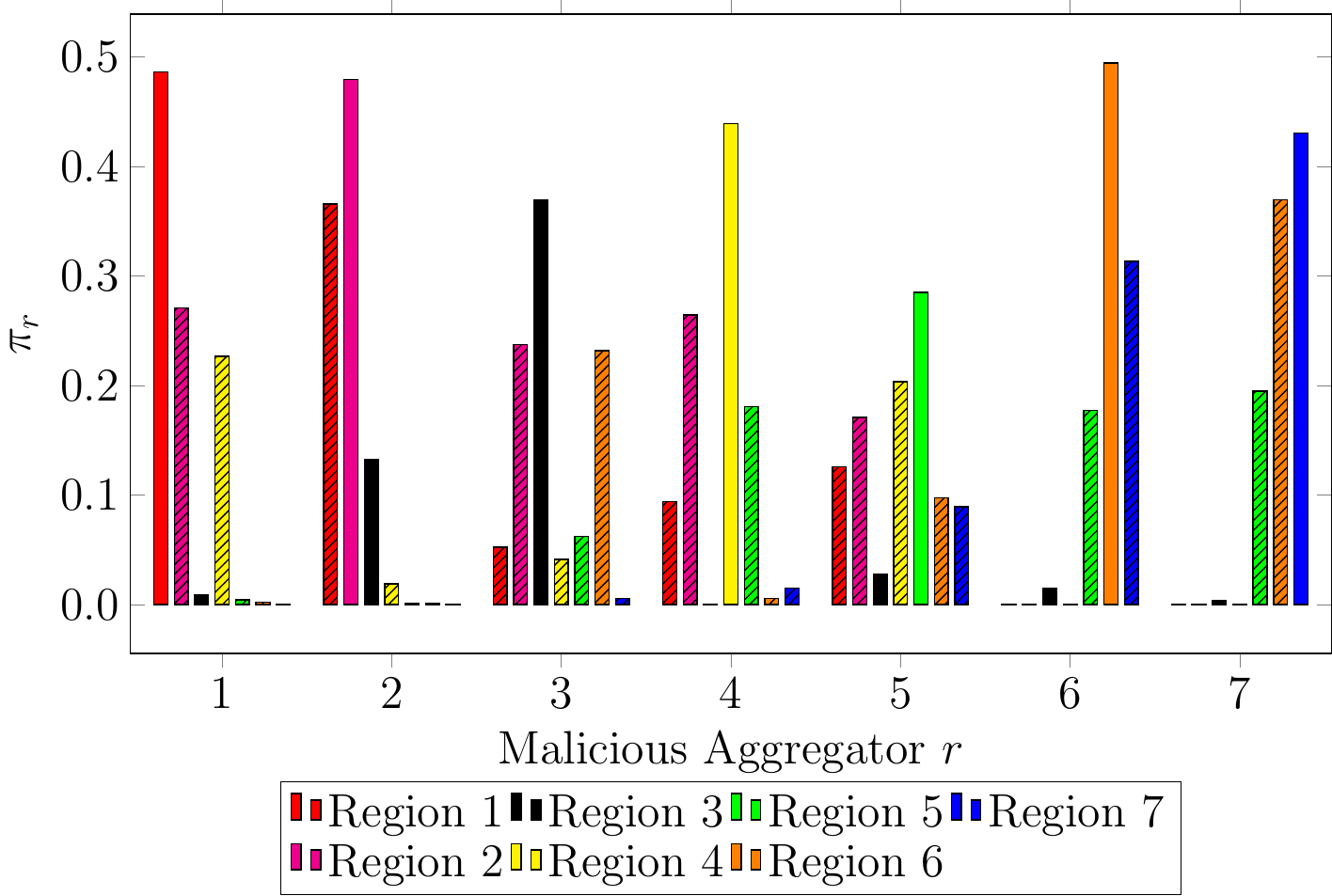}
    \caption{The stationary distribution, $\bm{\pi}$, of $\bm{B}$ under FDI attack. $\bm{\pi}$ represents the trust score: higher the value of $\pi_r$, lower is the trust in the aggregator $r$.}
    \label{fig:stat_dist}
\end{figure}
\par When aggregator $j$ is the attacker, the FDI attack constitutes an injection of an attack vector $\bm{a}^{(j)}$ to the communications received by the neighbors of the attacker in equation~\eqref{eq:ADMM_phi}. In each iteration of the algorithm, the attack vector is chosen randomly while satisfying $\|\bm{a}^{(j)}\|_2 = 0.5\sqrt{|\bm{S}_{ji}\bm{x}^{(j)}(t)|}$, where $|\bm{S}_{ji}\bm{x}^{(j)}(t)|$ is the length of the vector $\bm{S}_{ji}\bm{x}^{(j)}(t)$, i.e., the number of common variables shared by region $j$ and its neighbor $i$. Figure~\ref{fig:noattack} shows the convergence of the four common variables shared by aggregators $1$ and $2$ when there is no attack and in contrast, Figure~\ref{fig:attack}, shows their divergence when region $1$ is the attacker. As seen in Figure~\ref{fig:case141}, these four common variables correspond to the following buses: $42,43,54,$ and $73$. In Figure~\ref{fig:noattack}, both the aggregators converge to the optimal point $p^\star_b,~\forall b \in \{42,43,54,73\}$. The attack in Figure~\ref{fig:attack} is such that, the parameters do come to a consensus, but do not converge (i.e., they diverge). This will stop the verification algorithm from finishing. To stop such kinds of attacks from hampering our setup, we utilize the algorithm described in Section~\ref{sec:verification}.

As discussed in Section~\ref{sec:verification}, the stationary distribution of, $\bm{\pi}$, of the disagreement matrix, $\bm{B}$, is used to detect FDI attacks from selfish entities. Figure~\ref{fig:stat_dist} contains seven sets of bar plots, such that set $j$ represents the scenario where aggregator region $j$ was injecting false data into Algorithm~\ref{alg:state_estimation}. Each set contains seven bars representing the element of vector $\bm{\pi}$ corresponding to each of the seven aggregators. It is to be noted that in Figure~\ref{fig:stat_dist}, for an aggregator, as the height of the bar increases, it is seen as more untrustworthy by the other aggregators.
For example, in scenario $1$ where aggregator $1$ is the attacker, the corresponding bar plot indicates that the network of aggregators trust aggregator $1$ the least ($\pi_1$ is highest). Similarly, in each of the other sets $j\in [2,\ldots,7]$, we set aggregator $j$ as the attack and observe the level of trust the network places in the aggregators. As can be seen, in each scenario, the corresponding attacker amasses the lowest trust level (indicated by the highest $\pi_j$ value). It is interesting to note the distribution of distrust in the network. Consider scenario $1$ and notice that the distrust is spread among aggregators $1, 2$ and $4$. This can be explained by noticing the $\mathcal{G}_c$ graph shown in Figure~\ref{fig:case141}, where we see that aggregators $2$ and $4$ are neighbors of aggregator $1$. Set $5$ is another interesting scenario. Here, aggregator $5$ is the dishonest entity, but while it is seen as the most untrustworthy of the lot, it is not seen as untrustworthy as the attackers from the other scenarios. We also notice that the distrust is spread out the most here. The reason for this, referring to $\mathcal{G}_c$ again, might be because aggregator $5$ has one of the highest degrees and is the only \textit{cut vertex} (a vertex that when removed (with its boundary edges) from a graph creates more components than previously in the graph) in the network. This method may be used to identify FDI attackers.


\section{Conclusion}\label{sec:conclusion}
This paper discusses the lack in the resiliency of Blockchain based Transactive Energy frameworks to threats from within the network. The proposed Robust State Verification algorithm can output the actual state of the grid using the sensor measurements, and is well equipped to detect the presence of FDI attacks. The algorithm fits into the Blockchain Transactive Energy framework between the pricing and the billing phases, and may be used to hold all the players accountable. The distributed architecture of the optimization formulation is amenable to be solved via ADMM. The algorithm converges to the actual state of the grid in the absence of any attacks, and in the presence of random FDIs by the attacker to the ADMM states, will terminate with a convergence of the trust score vector using which the attacker is identified. Future work will include testing the algorithm in a Transactive Energy framework empowered by a permissioned Blockchain architecture.

\addtolength{\textheight}{-12cm}   


\bibliographystyle{IEEEtran}
\bibliography{nikref,thebib,blockchain}

\end{document}